\documentclass{article}
\usepackage{dcase2017,amsmath,graphicx,url,times,booktabs, tabularx}
\usepackage{flushend}
\usepackage{color,soul}

\usepackage{multirow}

\title{Sound event detection using weakly labeled dataset with \\ stacked convolutional and recurrent neural network}

\name{Sharath Adavanne, Tuomas Virtanen \thanks{The research leading to these results has received funding from the European Research Council under the European Union’s H2020 Framework Programme through ERC Grant Agreement 637422 EVERYSOUND. The authors also wish to acknowledge CSC-IT Center for Science, Finland, for computational resources.}}
\address{Department of Signal Processing , Tampere University of Technology}

\begin{document}

\ninept
\maketitle

\begin{sloppy}

\begin{abstract}
This paper proposes a neural network architecture and training scheme to learn the start and end time of sound events (strong labels) in an audio recording given just the list of sound events existing in the audio without time information (weak labels). We achieve this by using a stacked convolutional and recurrent neural network with two prediction layers in sequence one for the strong followed by the weak label. The network is trained using frame-wise log mel-band energy as the input audio feature, and weak labels provided in the dataset as labels for the weak label prediction layer. Strong labels are generated by replicating the weak labels as many number of times as the frames in the input audio feature, and used for strong label layer during training. We propose to control what the network learns from the weak and strong labels by different weighting for the loss computed in the two prediction layers. The proposed method is evaluated on a publicly available dataset of 155 hours with 17 sound event classes. The method achieves the best error rate of 0.84 for strong labels and F-score of 43.3\% for weak labels on the unseen test split. 
\end{abstract}


\section{Introduction}
\label{sec:intro}
Sound event detection (SED) is the task of recognizing sound events and its respective start and end timings in an audio recording. Recognizing such sound events and its temporal information can be useful in different applications such as surveillance~\cite{surveillance,surveillance_audio}, biodiversity monitoring~\cite{Karol2015, environmentalSED} and query based multimedia retrieval~\cite{contentRetrieval}. Traditionally, SED has been tackled with datasets that have temporal information for each of the sound event present~\cite{dcase2016task3web,Adavanne2017}. We refer to such temporal information of sound events as strong labels in this paper. 

The internet has a vast collection of audio data. Many collaborative and social websites like Freesound~\footnote[1]{https://freesound.org/} and YouTube~\footnote[2]{https://www.youtube.com/} allow users to upload multimedia with metadata like captions and tags. We can potentially automate the collection of audio data associated with a given tag from these online sources in considerably less time and manual effort. Recently, Gemekke et al.~\cite{Gemmeke2017} carried out this with 632 sound event tags on YouTube and collected nearly two million 10 second audio recordings. While these tags indicate that the sound event is present in the audio recording, the tags do not contain the information as to how many times they occur or at what time they occur. In this paper, we call such tags without any temporal information as weak labels. The task of identifying weak labels of an audio is also referred as audio tagging in literature~\cite{Xu2017, Adavanne2017_eusipco}.

Collecting and annotating data with strong labels to train SED methods is a time-consuming task involving a lot of manual labor. On the other hand, collecting weakly labeled data takes much less time to annotate manually, since the annotator has to mark only the active sound event classes and not its exact time boundaries. If we can build SED methods which can learn strong labels from such weakly labeled data, then the methods can learn on a large amount of data. In this paper, we propose to implement such a strong label learning SED method using weakly labeled training data.

Similar research of using weakly labeled data to learn strong labels has been done in neighboring audio domains such as music~\cite{Liu2016,Mandel2008}, and bird classification~\cite{Munoz2015,Briggs2012}. Liu et al.~\cite{Liu2016} used a fully convolutional neural network (FCN) to recognize instruments and tempo for each time frame of an audio clip given only the clip level information. They further extended this network to sound event detection~\cite{Su2017} and experimented on publicly available datasets. The advantage of using an FCN is it can handle audio input of any length. On the other hand, the limitation is that the frame-wise strong labels are obtained by an upscaling layer which replicates segment-wise output to as many number of frames required. Similar FCN as~\cite{Su2017} was proposed in~\cite{Kumar2017} without the upscaling layer, thereby estimating labels for short segments of length 1.5 s instead of frame wise labels. The study compares the performance of this FCN with a VGG-like network~\cite{Simonyan2015} like network which outputs sound event labels in segments of 1.5 s. The FCN network is trained using the entire audio, and its respective weak label. On the other hand, the VGG network is trained on sub-segments of the entire audio, assuming that the recording level weak label annotation remains the same in all its sub-segments. The study showed that using an FCN performs better SED than using the VGG method. Kumar et al.~\cite{Kumar2016} proposed a multiple instance learning (MIL) approach~\cite{Dietterich1997} for this task, though the results were promising the approach was claimed to be not scalable to large datasets by the same authors in~\cite{Kumar2017}.

Sound events in real life most often overlap each other. A SED method which can recognize such overlapping sound events is referred as polyphonic SED method. The state of the art for polyphonic SED, trained using strong labels, was proposed recently in~\cite{emre_TASLP2016}, where log mel-band energy feature was used along with a stacked convolutional and recurrent neural network and evaluated on multiple datasets. Similar stacked convolutional and recurrent neural network has also been shown to outperform state of the art methods in audio tagging tasks~\cite{Xu2017,Adavanne2017_eusipco}. Motivated by the performance of this method in SED and audio tagging, in this paper, we propose to extend the method to perform both SED and audio tagging together, given only the audio and its respective weak labels. In particular, we use the log mel-band audio feature extracted from the audio and extend the stacked convolutional and recurrent neural network to predict two outputs sequentially, the strong followed by the weak labels. To train the proposed network we generate dummy strong labels by replicating the weak labels as many times as the number of frames in the audio input feature. We further propose to control the information that the network learns by separately scaling the loss calculated in the weak and strong prediction layers.


Networks similar to the proposed stacked convolutional and neural network are the current state of the arts for audio tagging~\cite{Xu2017,Adavanne2017_eusipco}. This shows that the architecture is capable of learning the relevant information in temporal domain and mapping it to active classes. In this paper, we show that the proposed training scheme can extract this temporal information that the network is learning in the intermediate layers and can be used as strong labels. In comparison to previous works~\cite{Su2017, Kumar2017}, the proposed method supports higher time resolution for strong labels by its inherent design.

The feature extraction and the proposed network is described in Section~\ref{sec:method}. The dataset, metric and evaluation procedure is discussed in Section~\ref{sec:eval}. Finally, the results and discussions of the evaluation performed are presented in Section~\ref{sec:results}.

\begin{figure}[!ht]
  \centering
  \centerline{\includegraphics[width=10cm, height=15cm,keepaspectratio]{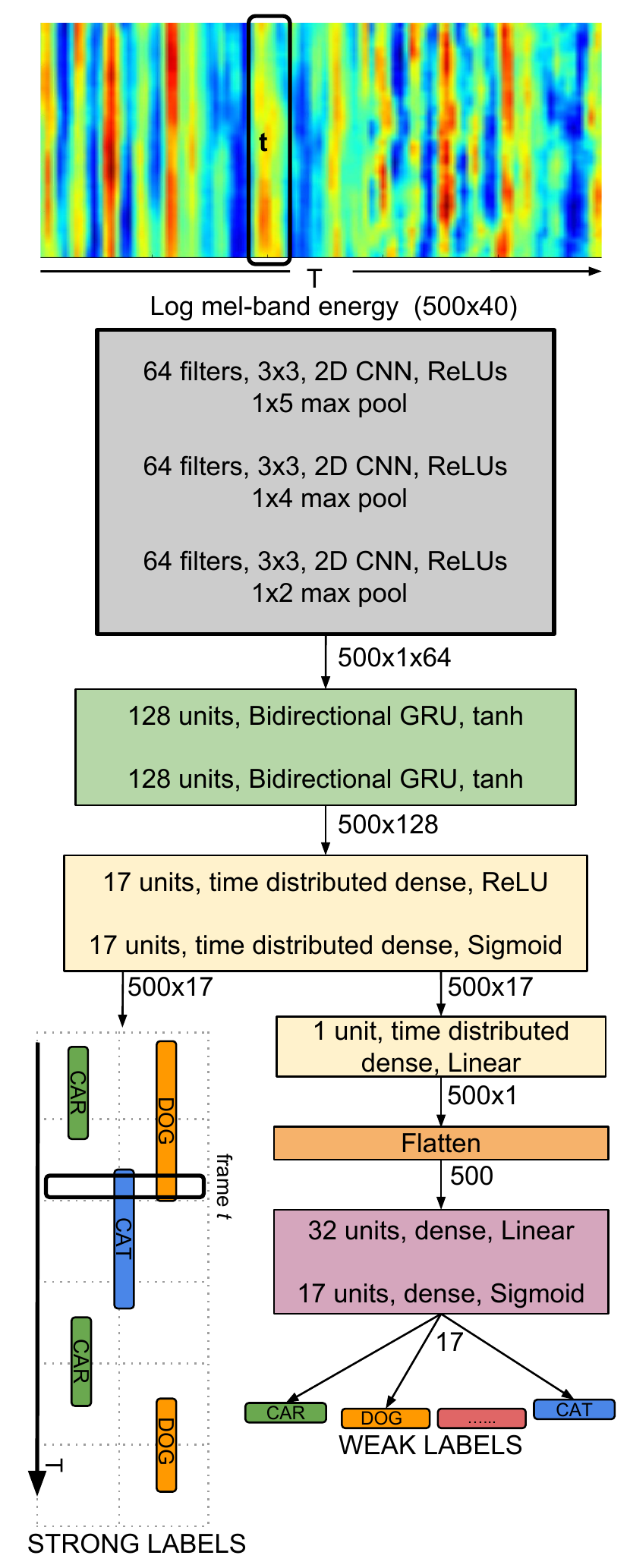}}
  \caption{Stacked convolutional and recurrent neural network for learning strong labels from weak labels.}\vspace{-10pt}
  \label{fig:network}
\end{figure}

\section{Method}
\label{sec:method}
Figure~\ref{fig:network} shows the overall block diagram of the proposed method. The log mel-band energy feature extracted from the audio is fed to a stacked convolutional and recurrent neural network, which sequentially produces the strong labels followed by the weak labels.

Audio features are calculated using overlapping windows on the input audio of length 10-seconds, resulting in $T$ frames of the feature. The proposed neural network maps these features into strong labels first, and further, the strong labels are mapped to weak labels. For an input of $T$ frames, and a total number of sound classes $C$ in the dataset, the network predicts $C$ for each of the $T$ time frames as strong label output and just $C$ as weak label output. The predicted outputs for each of the sound class is in the continuous range of [0, 1], where one signifies the presence of the sound class and zero the absence. The details of the feature extraction and the network are presented below.

\vspace{-5pt}
\subsection{Feature extraction}
\label{ssec:feat}
Log mel-band energy ($mbe$) is extracted in 40 ms Hamming windows with 50\% overlap. In total 40 mel bands are used in the 0-22050 Hz range. For a given 10 second audio input, the feature extraction block produces a 500$\times$40 output ($T$ = 500).

\vspace{-5pt}
\subsection{Neural network}
\label{ssec:dnn}
The input to the proposed network is the $T\times40$ $mbe$ feature as shown in Figure~\ref{fig:network}. The local shift-invariant features of this input are learned using CNN layers in the beginning. We use a $3\times3$ receptive field and pad the output with zeros to keep the size same as input in all our CNN layers. The max-pooling operation is performed along the frequency axis after every layer of CNN to reduce the dimension to $T\times1\times N$, where $N$ is equal to the number of filters in the last CNN layer. We do not perform max-pooling along the time axis to preserve the input time resolution. The CNN layers activation is further fed to a bi-directional gated recurrent units (GRU) with tanh activation to learn the long-term temporal structure of sound events, followed by time distributed fully-connected (dense) layers to reduce the feature-length dimensionality. The time resolution of $T$ frames is unaltered in both the GRU and dense layers. Since we have to predict multiple labels simultaneously, we use sigmoid activation in the last dense layer. This prediction layer outputs the strong labels present in the input audio, and we refer to this as \textit{strong output} in future. The dimensions of the strong labels are $T\times C$. We calculate the strong label loss on this output. Further, we reduce the activation dimensionality and remove the framing information using dense layers and map it to the $C$ weak labels present in the audio. We refer to this weak label prediction layer as \textit{weak output} in future and calculate the weak label loss on its output. The total loss of the network is then calculated as the weighted sum of strong and weak losses.

During the training, the loss at weak and strong outputs was weighed differently to facilitate learning from one output more than the other. In other words, during training, the weak labels along with the weighting scheme help control the learning of strong labels. On the other hand, during testing, the weak labels are obtained from the predicted strong labels.

Batch normalization~\cite{batchNorm} is performed on the activations of every CNN layer. We train the network for 1000 epochs using binary cross-entropy loss function for both the strong and weak outputs, and Adam~\cite{adamKeras} optimizer. Early stopping was used to reduce the overfitting of the network to training data. The training was stopped if the sum of the error rate of strong labels and F-score of weak labels (see Section~\ref{ssec:metric}) referred as the training metric in future did not improve for more than 100 epochs. We used dropout\cite{Dropout} after every layer of the network as a regularizer to make the training generic and work on unseen data. The implementation of the network was done using Keras~\cite{chollet2015keras} with Theano~\cite{theano} as backend.

\section{Evaluation} 
\label{sec:eval}
\subsection{Dataset}
\label{ssec:data}
The method is evaluated using a subset of the recently released Audioset data by Google~\cite{Gemmeke2017}. This subset was organized as part of a challenge in the Detection and Classification of Acoustic Scenes and Events (DCASE)~\cite{dcase2017task4}.

The dataset consists of a training, testing and evaluation split. The training split consists of 51,172 recordings, and the testing split consists of 488 recordings. All recordings are of 10-second length, monochannel and sampled at 44100 Hz. All these recordings have been collected from publicly uploaded Youtube videos as explained in~\cite{Gemmeke2017}. Different methods trained on this training and testing split were benchmarked using the unseen evaluation split of 1103 recordings at the DCASE 2017 challenge~\cite{dcase2017task4}.

The dataset contains 17 labels in total and each recording can have more than one label. Strong labels are provided only for the testing split, while weak labels are provided for both the splits. In order to train our network, we need strong labels in the training data as well. We generate this by replicating the weak labels for every time frame of the audio and use them as strong labels.


\subsection{Metric}
\label{ssec:metric}
We evaluate our method in a similar fashion as the challenge~\cite{dcase2017task4}. Evaluation are performed individually on the weak and strong label predictions.

The weak labels are evaluated by calculating the total number of recalls ($R$), its respective precision ($P$) and the F-score as $R = TP\, /\, (TP+FN)$, $P = TP\,/\, (TP+FP)$ and $F = 2\cdot P \cdot R\, /\, (P + R)$ respectively. Where, true positives ($TP$) is the number of times the method correctly predicted the ground-truth label. False positives ($FP$) is the number of times the method predicted incorrectly the ground-truth labels. False negative ($FN$) is the number of times the method did not predict a ground-truth label.

The strong labels are evaluated using a segment based F-score and error rate (ER) as proposed in~\cite{metrics}. According to which the F-score is calculated as
\begin{equation}
F = \frac{2 \cdot \sum_{k=1}^{K} TP(k)}{2 \cdot \sum_{k=1}^{K}TP(k)+ \sum_{k=1}^{K}FP(k)+ \sum_{k=1}^{K}FN(k)},
\end{equation}
where $TP(k)$, $FP(k)$ and $FN(k)$ are the true positives, false positives and false negatives respectively calculated for each of the $K$ segments.
The ER is calculated as
\begin{align}
ER = \frac{\sum_{k=1}^{K} S(k) + \sum_{k=1}^{K} D(k) + \sum_{k=1}^{K} I(k)}{\sum_{k=1}^{K} N(k)},
\end{align}
where $N(k)$ is the total number of labels active in a given segment $k$. $S(k)$, $D(k)$ and $I(k)$ are the substitutions, deletions and insertions respectively measured for each of the $K$ segments as $S(k) = \min(FN(k),\, FP(k))$, $D(k) = \max(0,\, FN(k)-FP(k))$ and $I(k) = \max(0,\, FP(k)-FN(k))$.

We use a segment length of one second for our strong label metrics. The ideal F-score is 100 and ER is zero. 

\vspace{-5pt}
\subsection{Baseline}
\label{ssec:baseline}
The baseline method for the dataset is provided by~\cite{dcase2017task4}. It is a basic method to provide a comparison point for other methods using the dataset. This baseline method uses $mbe$ as the audio feature. The network used is a fully-connected one with two hidden layers, each with 50 units and 20\% dropout, followed by a prediction layer with as many sigmoid units as the number of classes in the dataset. A context of five frames of the audio feature is used for training the network along with binary cross-entropy loss and Adam optimizer. The evaluation metric scores for the baseline method are shown in Table~\ref{Table:dropout}. The network is trained by replicating the weak labels as many number of times as the number of frames in the input audio feature. During testing, the weak labels are obtained outside the network by identifying the sound events active in the strong labels.

\vspace{-5pt}
\subsection{Evaluation procedure}
\label{ssec:eval_proc}
The stacked convolutional and recurrent neural network is trained with $mbe$ as input, the weak labels provided in the dataset as weak output and the strong labels generated by replicating weak labels for each time frame as strong output. 

Given that the data is huge and the hardware has memory constraints, the training time can be long (about 1800 s/epoch on our hardware). We cannot perform an extensive hyperparameter search in the limited time, hence we start with a similar network configuration as in~\cite{Adavanne2017}, and perform a random search~\cite{Bergstra2012} by varying the number of units/filters in each of the layers until no under or over-fitting is observed while having a strong training metric. Since the dataset is large and is uploaded by different users, we assume that there will be enough variability and hence do not use any regularizer. The best configuration with highest training metric is as shown in Figure~\ref{fig:network}. This configuration has around 218,000 parameters. Other configurations with higher number of parameters, up to 2,000,000, did not show any substantial improvement over the chosen configuration.

On finalizing the network, in order to be sure of our assumption that the high variability in data will not result in an over-fitting model, we experiment using dropout for each layer in our network as a regularizer and vary it in the set of \{0.05, 0.15, 0.25, 0.5 and 0.75\}. We use the same dropout for all layers in this study.

The weights for the two prediction layers were experimented with different combinations in the logarithmic set of \{0.002, 0.02, 0.2, 1\}. The number 0.002 is motivated from the ratio of the total number of time frames for the weak label (1) to the number of frames for the strong label (500). 

\begin{table}[t]
\centering
\begin{tabular}{l|ccc|cc}
 & \multicolumn{3}{c|}{Weak labels} & \multicolumn{2}{c}{Strong labels} \\ \cline{2-6}
Dropout & P & R & F & ER & F \\ \hline
Baseline~\cite{dcase2017task4} & 12.2 & 14.1 & 13.1 & 1.02 & 13.8 \\ \hline 
0.05 & 44.6 & 37.8 & 40.9 & 0.86 & 38.6 \\
\bf{0.15} & \bf{47.5} & \bf{39.7} & \bf{43.3} & \bf{0.84} & \bf{38.8} \\
0.25 & 43.0 & 35.0 & 38.6 & 0.86 & 33.8 \\
0.5 & 21.5 & 17.3 & 19.2 & 0.99 & 8.6 \\
0.75 & 12.3 & 9.9 & 11.0 & 1.15 & 8.0
\end{tabular}
\caption{Evaluation metric scores for weak and strong labels for different dropout values.}\vspace{-15pt}
\label{Table:dropout}
\end{table}

\section{Results and discussion} 
\label{sec:results}

The evaluation metric scores for weak and strong labels for different dropout values tried are compared in Table~\ref{Table:dropout}. The best training metric of 43.3\% F-score for weak labels and 0.84 ER for strong labels was achieved with 0.15 dropout for the proposed network. This study was performed by having the same weight of one for weak and strong outputs during training. In comparison with the baseline~\cite{dcase2017task4} score of 1.02 ER for strong labels and 13.1\% F-score for weak labels, this is a significant improvement. 

We used the above-estimated dropout of 0.15 and studied how the network learns when provided with different weights for weak and strong outputs and present the results in Table~\ref{T:loss_weight_compare}. For example, from the first row of the table, the loss at strong output was scaled with 0.002 while the loss at weak output was unscaled. Since the strong labels for training were generated by replicating the weak labels, they are not the actual true labels, hence by intuition, we assumed higher weighting for weak labels will give better training metric. From experimentation, it was seen that the best training metric was actually obtained by using the same weight of one for both the weak and strong outputs. Another interesting observation is that the ER and F-score for strong labels improve when strong output is given more weight than the weak output, even though the strong labels used while training is `weak' in the sense of correctness. On the other hand, this also results in poor metric for weak labels.

We analyzed the predicted labels of our configuration with an equal weight of one for weak and strong outputs which achieved the best training metric. Among the weak labels, the vehicular sound events - train and skateboard, warning events - fire engine siren and civil defense siren were seen to have the highest F-scores of over 60\%. On the other hand, sound events - ambulance siren, car alarm, car passing, reverse beeps, train horn had zero F-score. The same sound events and the trend were observed for strong labels. 

In order to understand what our method is learning, we visualized the activations in the first convolutional layer of the network for a given output class. The visualizations are done using the saliency map~\cite{Simonyan2013} approach implementation in keras-vis~\cite{keras_vis}. The saliency map is the gradient of output class with respect to the input feature. An example of such visualization is shown in Figure~\ref{fig:vis} for the recording `--jc0NAxK8M\_30.000\_40.000' in the test dataset. The top and center sub-plots are the activations of the first convolutional layer for the strong and weak output of sound class `car'. The bottom subplot shows the ground truth marked in red dotted line over the input $mbe$ feature. We see from the activation map of both strong and weak heat maps that the network is actually learning the sound event from a relevant time period in the $mbe$ feature.

\begin{table}[t]
\centering
\resizebox{\columnwidth}{!}{\begin{tabular}{l|l|ccc|c|cc|cc}
\multicolumn{1}{c|}{\multirow{2}{*}{\begin{tabular}[c]{@{}c@{}}Strong\\ Weight\end{tabular}}} & \multicolumn{1}{c|}{\multirow{2}{*}{\begin{tabular}[c]{@{}c@{}}Weak \\ Weight\end{tabular}}} & \multicolumn{4}{c|}{Weak labels} & \multicolumn{4}{c}{Strong labels} \\ \cline{3-10}

\multicolumn{1}{c|}{} & \multicolumn{1}{c|}{} & P & R & F & F\textsubscript{ch} & ER &  F  & ER\textsubscript{ch} & F\textsubscript{ch} \\ \hline
0.002 & 1 & 44.9 & 37.0 & 40.5 & & 1.38 &10.9 &  & \\
0.02 & 1 & 44.2 & 36.5 & 40.0 & & 1.13 & 17.0 &  & \\
0.2 & 1 & \bf{47.5} & 39.6 & 43.2 & \bf{46.6} & 0.84 & 38.1 & 0.80 & 48.3\\
1 & 1 & \bf{47.5} & \bf{39.7} & \bf{43.3} & 45.5 & 0.84 & 38.8 & 0.81 & 47.9 \\
1 & 0.2 & 47.3 & 39.5 & 43.0 & 44.5 & 0.84 & 38.6 & 0.82 & 48.9 \\
1 & 0.02 & 25.5 & 20.6 & 22.8 & & \bf{0.81} & 41.1 & &  \\
1 & 0.002 & 20.5 & 16.5 & 18.3 & 26.3 & \bf{0.81} & \bf{42.4} & \bf{0.79} & \bf{49.0}
\end{tabular}}
\caption{Evaluation metric scores for different combinations of weights for strong and weak label loss. The scores with subscript \textsubscript{ch} represents the challenge results.}\vspace{-10pt}
\label{T:loss_weight_compare}
\end{table}

\begin{figure}[!ht]
  \centering
  \centerline{\includegraphics[width=\columnwidth,keepaspectratio]{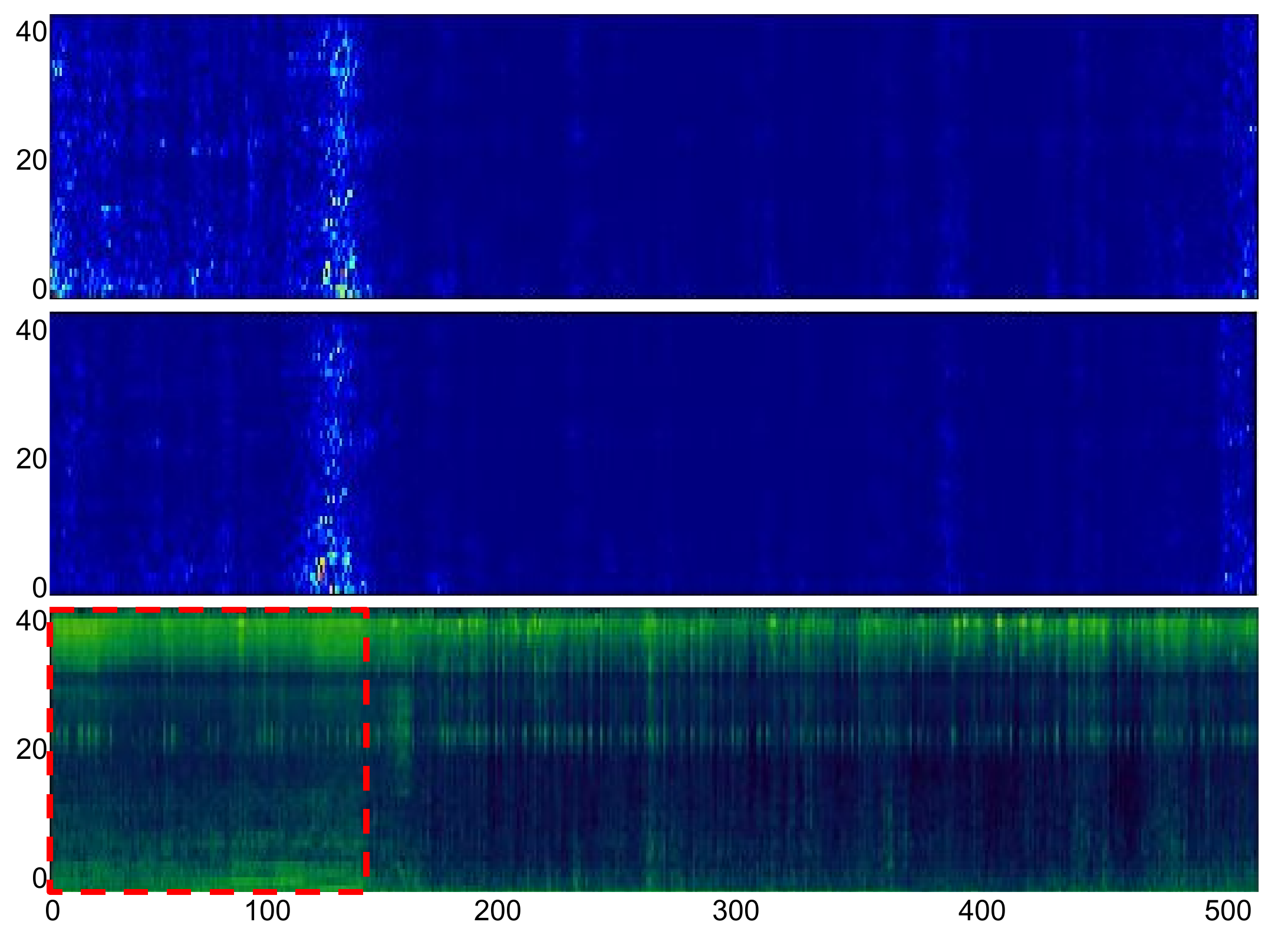}}
  \caption{Visualization of the activations in the first layer of CNN for strong (top) and weak (center) prediction of sound event class `car' in `--jc0NAxK8M\_30.000 \_40.000' recording. The bottom plot shows the input to the network, the log mel-band feature of the recording, where the sound event class is active in the region bounded by the red dotted line.}\vspace{-10pt}
\label{fig:vis}
\end{figure}

\subsection{DCASE 2017 challenge results}
The results of the proposed method on the evaluation split of DCASE 2017 challenge~\cite{dcase2017task4} is presented in Table~\ref{T:loss_weight_compare}. Four systems with different strong and weak output weighting were chosen based on their performance on test split. Similar trend of better strong label score when strong output is weighed more was observed on evaluation data (ER\textsubscript{ch} = 0.79) as well. In comparison,~\cite{Xu2017_dcase} obtained the best weak label F-score of 55.6\% and~\cite{Lee2017} obtained the best strong label error rate of 0.66.
\vspace{-10pt}
\section{Conclusion} 
\label{sec:conclusion}
The task of learning temporal information of sound events in an audio recording, given only the sound events existing in the audio without the time information is tackled in this paper. A stacked convolutional and recurrent neural network architecture with two prediction layer outputs and a training scheme was proposed in this regard. This network was trained using different weights for the loss calculated in the two prediction layers. Even though the strong labels used for training were just repeated weak labels, it was observed that the network learned the relevant strong labels correctly when the weighting for the two prediction layers was equal. This evaluation was carried out on a publicly available dataset of 155 hours duration. An error rate of 0.84 for strong labels and F-score of 43.3\% for weak labels was achieved on the test data.

\bibliographystyle{IEEEtran}
\bibliography{refs}
\end{sloppy}
\end{document}